\documentclass[graybox]{svmult}

\usepackage{helvet}         
\usepackage{courier}        
\usepackage{type1cm}        
\usepackage{makeidx}         
\usepackage{graphicx}        
\usepackage{multicol}        
\usepackage[bottom]{footmisc}

\usepackage{amsmath,latexsym,amssymb}
\usepackage{young}  

\def\nn{\nonumber}
\def\qdots{\mathinner{\mkern1mu\raise1pt\vbox{\kern7pt\hbox{.}}\mkern2mu
 \raise4pt\hbox{.}\mkern2mu\raise7pt\hbox{.}\mkern1mu}}

\def\g{\mathfrak{g}}
\def\gl{\mathfrak{gl}}
\def\ssl{\mathfrak{sl}}

\def\so{\mathfrak{so}}
\def\osp{\mathfrak{osp}}
\def\sp{\mathfrak{sp}}

\DeclareMathOperator{\sdim}{sdim}

\DeclareMathOperator{\ch}{char}
\def\qdots{\mathinner{\mkern1mu\raise1pt\vbox{\kern7pt\hbox{.}}\mkern2mu \raise4pt\hbox{.}\mkern2mu\raise7pt\hbox{.}\mkern1mu}}

\begin{document}

\title*{On superdimensions of some infinite-dimensional irreducible representations of $\osp(m|n)$}
\titlerunning{Superdimensions of $\osp(m|n)$ irreps}
\author{N.I.~Stoilova, J.~Thierry-Mieg and J.~Van der Jeugt}
\institute{
N.I.\ Stoilova \at Institute for Nuclear Research and Nuclear Energy,
Boul.\ Tsarigradsko Chaussee 72, 1784 Sofia, Bulgaria, \email{stoilova@inrne.bas.bg} 
\and 
J.\ Thierry-Mieg \at NCBI, National Library of Medicine, National Institute of Health,
8600 Rockville Pike, Bethesda MD20894, USA, \email{mieg@ncbi.nlm.nih.gov}
\and
J.\ Van der Jeugt \at Department of Applied Mathematics, Computer Science and Statistics, Ghent University,
Krijgslaan 281-S9, B-9000 Gent, Belgium, \email{Joris.VanderJeugt@UGent.be}
}
\maketitle

\abstract{
In a recent paper characters and superdimension formulas were investigated for the class of representations
with Dynkin labels $[0,\ldots,0,p]$ of the Lie superalgebra $\osp(m|n)$.
Such representations are infinite-dimensional, and of relevance in supergravity theories provided their 
superdimension is finite. 
We have shown that the superdimension of such representations coincides with the dimension of a $\so(m-n)$ representation.
In the present contribution, we investigate how this $\osp(m|n)\sim\so(m-n)$ correspondence can be extended
to the class of $\osp(2m|2n)$ representations with Dynkin labels $[0,\ldots,0,q,p]$.
}

\section{Introduction}
\label{sec:1}

Chiral spinors and self dual tensors of the Lie superalgebra $\osp(m|n)$ play a prominent role in some models of supergravity theory~\cite{Baulieu,Preitschopf}.
As representations, these spinors and self dual tensors are characterized by Dynkin labels $[0,\ldots,0,p]$, where $p=1$ for the chiral spinor
and $p=2$ for the self dual tensor. 
It will be interesting to consider the class of representations with arbitrary positive integer~$p$.
Although all Dynkin labels are nonnegative integers, the corresponding representations are infinite-dimensional
(as they do not satisfy the extra condition in Kac's list of finite-dimensional irreducible representations~\cite{Kac,Kac1}).
In~\cite{STV}, we showed that the superdimension of these representations coincides with the dimension of the corresponding
$\so(m-n)$ representation.
Herein, the algebra should be interpreted differently when $m-n$ is negative: as $\sp(n-m)$ when $n-m$ is even, 
and as $\osp(1|n-m-1)$ when $n-m$ is odd.

The results of~\cite{STV} rely on the knowledge of the character for such $\osp(m|n)$ representations.
In particular, the expansion or formulation of this character in terms of supersymmetric Schur functions 
turned out to be the crucial ingredient in order to obtain the $\osp(m|n)\sim\so(m-n)$ correspondence.

In the present paper, we shift our attention to the class of $\osp(2m|2n)$ representations with Dynkin labels $[0,\ldots,0,q,p]$.
In the distinguished Dynkin diagram of $\osp(2m|2n)$, all nodes have zero labels and only the two nodes of the fork have a 
non-negative integer label. 
Such representations are again infinite-dimensional.
Our idea to deal with these representations is as follows:
we will first investigate the finite-dimensional $\so(2k)$ representations of type $[0,\ldots,0,q,p]$,
conjecture that the $\osp(m|n)\sim\so(m-n)$ correspondence still holds, and as such obtain interesting new characters
of $\osp(2m|2n)$ representations.

\section{Preliminaries and definitions}
\label{sec:2}

The character formulas used in this paper are expressed in terms of symmetric or supersymmetric Schur functions, which are labelled by partitions. 
So it will be useful to recall some notation for this. The standard reference is~\cite{Mac}.
A partition $\lambda=(\lambda_1,\lambda_2,\ldots,\lambda_n)$ of weight $|\lambda|$ and length $\ell(\lambda)\leq n$
is a sequence of non-negative integers satisfying the condition $\lambda_1\geq\lambda_2\geq\cdots\geq\lambda_n$, such that their
sum is $|\lambda|$, and $\lambda_i>0$ if and only if $i\leq \ell(\lambda)$. 
It is common to represent (and sometimes identify) a partition by its Young diagram. 
For example, the Young diagram of $\lambda=(6,4,4,2)$ is given by the first figure in~\eqref{youngdiag}.
\begin{equation}
\begin{Young}
&&&&&\cr
&&&\cr
&&&\cr
&\cr
\end{Young} \qquad\qquad
\begin{Young}
&&&&$\times$&$\times$\cr
&&&\cr
&&&$\times$\cr
&$\times$\cr
\end{Young}
\label{youngdiag}
\end{equation}
The conjugate partition $\lambda'$ corresponds to the Young diagram of $\lambda$ reflected about the main diagonal. 
For the above example, $\lambda'=(4,4,3,3,1,1)$. 
If $\lambda,\mu$ are two partitions, one writes $\lambda \supset \mu$ if the diagram of $\lambda$ contains that of $\mu$.
The difference $\lambda - \mu$ is called a skew diagram~\cite{Mac}. 
For example, if $\mu=(4,4,3,1)$, then the boxes of the skew diagram $\lambda - \mu$ are crossed in the second picture of~\eqref{youngdiag}.
A skew diagram is a {\em horizontal strip} if it has at most one box in each column. 
The number of boxes of the horizontal strip is its length.
The above example is a horizontal strip of length~4.

Partitions are used to label symmetric and supersymmetric functions. 
When dealing with characters of Lie algebras or Lie superalgebras, the Schur functions~\cite{Mac} or $S$-functions 
are the most useful basis. 
In terms of a set of $n$ independent variables $x=(x_1,x_2,\ldots,x_n)$, the Schur function $s_\lambda(x)$ (with $\lambda$ a partition) is a symmetric 
polynomial that can be defined by means of determinants~\cite{Mac}.
When dealing with two sets of variables $x=(x_1,\ldots,x_m)$ and $y=(y_1,\ldots,y_n)$, one can define the so-called supersymmetric Schur function
$s_\lambda(x|y)$~\cite{Berele,King1983}.
Here, $s_\lambda(x|y)$ is zero whenever $\lambda_{m+1}>n$. 
Following this, it is common to denote by ${\cal H}_{m,n}$ the set of all partitions with $\lambda_{m+1}\leq n$, 
i.e.\ the partitions (with their Young diagram) inside the $(m,n)$-hook. 

For characters of simple Lie algebras, ordinary Schur functions play a prominent role. 
Characters of finite-dimensional irreducible representation (irreps) of $\gl(n)$ or $\ssl(n)$ are directly given by a Schur function,
and characters of irreps of other simple Lie algebras can be expanded in Schur functions~\cite{KW}.
An irrep of $\mathfrak{gl}(n)$ is characterized by a partition $\lambda$ with $\ell(\lambda)\leq n$. 
In terms of the standard basis $\epsilon_1,\ldots,\epsilon_n$ of the weight space of $\mathfrak{gl}(n)$, the highest weight of this
representation is $\sum_{i=1}^n \lambda_i \epsilon_i$, and the representation space will be denoted by $V_{\mathfrak{gl}(n)}^{\lambda}$. 
Its character is given by $\ch V_{\mathfrak{gl}(n)}^{\lambda}  = s_\lambda(x)$, where $x_i=\hbox{e}^{\epsilon_i}$.

For Lie superalgebras, this role is played by the supersymmetric Schur functions, at least for certain classes of representations.
For a partition $\lambda\in{\cal H}_{m,n}$, the corresponding covariant representation of the Lie superalgebra $\mathfrak{gl}(m|n)$
will be denoted by $V_{\mathfrak{gl}(m|n)}^{\lambda}$.
In terms of the standard basis $\epsilon_1,\ldots,\epsilon_m,\ \delta_1,\ldots,\delta_n$ of the weight space of $\mathfrak{gl}(m|n)$, 
the highest weight of this representation is $\sum_{i=1}^m \lambda_i \epsilon_i + \sum_{j=1}^n \max(\lambda_j'-m,0)\delta_j$,
and the main result of~\cite{Berele} is
\begin{equation}
\ch V_{\mathfrak{gl}(m|n)}^{\lambda} = s_\lambda(x|y),
\end{equation}
where $x_i=\hbox{e}^{\epsilon_i}$ and $y_j=\hbox{e}^{\delta_j}$.

\section{Dimension, superdimension and $t$-dimension}
\label{sec:3}

As is well known, the character of a representation gives all information on the weight structure of the representation.
Sometimes, it is useful to consider certain {\em specializations} of characters, because of specific information that is needed, 
or because of elegant formulas that hold for certain specializations.
Let $V$ be a highest weight representation of a simple Lie algebra or Lie superalgebra, with highest weight $\Lambda$ and character $\ch V$. 
A well known specialization of the character of $V$ is the so-called $q$-dimension~\cite[Chapter~10]{Kac-book}.
The $q$-dimension of $V$ is nothing else than the specialization
\begin{equation}
\dim_q(V)=F(\hbox{e}^{-\Lambda} \ch V), \qquad\hbox{where}\qquad F(\hbox{e}^{-\alpha_i})=q,
\end{equation}
and the $\alpha_i$'s are the simple roots of the Lie (super)algebra. 
So this corresponds to the principal gradation of the Lie (super)algebra,
and one counts the dimension of the ``levels'' of the representation space starting from the top level (corresponding to the highest weight)
according to this gradation.

Here, we will be dealing with a different specialization, referred to as the $t$-dimension.
For a (simple) Lie algebra, of which the simple roots are commonly expressed in terms of the standard basis $\epsilon_1,\ldots,\epsilon_n$, 
one defines
\begin{equation}
\dim_t(V)=F_0(\hbox{e}^{-\Lambda} \ch V), \quad\hbox{where}\quad F_0 (\hbox{e}^{-\epsilon_i})=t.
\end{equation}
For a Lie superalgebra of type $\ssl$, $\gl$ or $\osp$, 
of which the simple roots are commonly expressed in terms of the standard basis $\epsilon_1,\ldots,\epsilon_m$,
$\delta_1,\ldots,\delta_n$, we define the $t$-dimension and the $t$-superdimension:
\begin{align}
\dim_t(V)&=F_0(\hbox{e}^{-\Lambda} \ch V), \quad\hbox{where}\quad F_0 (\hbox{e}^{-\epsilon_i})=t\hbox{ and }F_0(\hbox{e}^{-\delta_i})=t;\\
\sdim_t(V)&=F_1(\hbox{e}^{-\Lambda} \ch V), \quad\hbox{where}\quad F_1 (\hbox{e}^{-\epsilon_i})=t\hbox{ and }F_1(\hbox{e}^{-\delta_i})=-t.
\end{align}
Intuitively, the $t$-dimension again counts the dimension of levels of a representation starting from the top level, 
but according to a gradation different from the principal one.
Similarly, the $t$-superdimension counts the dimension of the same levels, but with alternating signs.
For finite-dimensional representations, putting $t=1$ in $\dim_t(V)$ gives the dimension of $V$, and putting $t=1$ in $\sdim_t(V)$ gives
its so-called superdimension (i.e.\ $\dim V_{\bar 0} - \dim V_{\bar 1}$, when $V=V_{\bar 0}\oplus V_{\bar 1}$ is written 
as the direct sum of its even and odd subspace).

Let us consider some examples.
For the orthogonal Lie algebra $\mathfrak{so}(2n+1)$, with simple roots $\epsilon_1-\epsilon_2, \ldots, \epsilon_{n-1}-\epsilon_n, \epsilon_n$,
we will focus on representations $V$ with Dynkin labels $[0,\ldots,0,p]$, for which the highest weight is $(\frac{p}{2}, \ldots, \frac{p}{2})$ in
the $\epsilon$-basis. 
For this representation, the character is~\cite{parafermion, BG1}
\begin{equation}
\ch [0,\ldots,0,p]_{\mathfrak{so}(2n+1)} 
= (x_1\cdots x_n)^{-p/2} \sum_{\lambda_1\leq p,\; \ell(\lambda)\leq n} s_\lambda (x).
\label{e4}
\end{equation}
So the sum is over all partitions $\lambda$ such that the Young diagram of $\lambda$ fits inside the $n\times p$ rectangle, of width $p$ and height $n$.
Specializing this character according to $F_0$, one finds:
\begin{equation}
\dim_t [0,\ldots,0,p]_{\mathfrak{so}(2n+1)} 
= \sum_{\lambda_1\leq p,\; \ell(\lambda)\leq n} \dim V^\lambda_{\mathfrak{gl}(n)} t^{|\lambda|}.
\label{e5}
\end{equation}
When the character is expressed in terms of Schur functions, as in~\eqref{e4}, 
it yields in fact the branching of the representation 
according to $\mathfrak{so}(2n+1) \supset \mathfrak{gl}(n)$.
When the character is specialized as in~\eqref{e5}, it is a polynomial in $t$ (or, in case of an infinite-dimensional representation, 
a formal power series in $t$) such that the coefficient of $t^k$ counts the dimension ``at level $k$'' according to the $\mathbb{Z}$-gradation
induced by the $\mathfrak{gl}(n)$ subalgebra of $\mathfrak{so}(2n+1)$.
For example, for $\so(7)$, one has
\begin{align*}
\dim_t [0,0,1]_{\mathfrak{so}(7)} &= 1+3t+3t^2+t^3,\\
\dim_t [0,0,2]_{\mathfrak{so}(7)} &= 1+3t+9t^2+9t^3+9t^4+3t^5+t^6,\\
\dim_t [0,0,2]_{\mathfrak{so}(7)} &= 1+3t+9t^2+19t^3+24t^4+24t^5+19t^6+9t^7+3t^8+t^9.
\end{align*}
The $q$-dimension, on the other hand, is a character specialization with a very different nature.
It is a character specialization closely related to Weyl's dimension formula, for which an explicit formula exists~\cite[(10.10.1)]{Kac-book}.
For the representations considered in this example, this yields (replacing $q$ by $q^2$ in order to avoid half-integer powers):
\begin{equation}
\dim_{q^2} [0,0,p]_{\mathfrak{so}(7)} = \frac{(1-q^{p+5})(1-q^{p+4})(1-q^{p+3})^2(1-q^{p+2})(1-q^{p+1})}{(1-q^{5})(1-q^{4})(1-q^{3})^2(1-q^{2})(1-q)}.
\end{equation}
So the $q$-dimension is a character specialization for the principal gradation of a Lie (super)algebra, 
leading to classical formulas.
The $t$-dimension is a character specialization related to the gradation coming from the $\gl(n)$ subalgebra (or $\gl(m|n)$ subalgebra),
thus typically related to the branching $\g \supset \gl(n)$ or $\g \supset \gl(m|n)$.

As a second example, let us consider the $t$-dimension for a class of representations of $\mathfrak{g}=\mathfrak{osp}(1|2n)$.
The notation is as follows~\cite{dict,Kac,Kac1}:
$\delta_j$ are the basis elements for the weight space of $\mathfrak{osp}(1|2n)$; 
the odd roots are given by $\pm\delta_j$ ($j=1,\ldots,n$), the even roots by $\delta_i-\delta_j$ ($i\ne j$) and $\pm(\delta_i+\delta_j)$, and the simple roots by
$\delta_1-\delta_2,\  \delta_2-\delta_3, \ldots, \delta_{n-1}-\delta_n,\ \delta_n$.
The subalgebra $\gl(n)$ is spanned by the root vectors corresponding to $\delta_i-\delta_j$.
The embedding $\gl(n)\subset\osp(1|2n)$ leads to a $\mathbb{Z}$-gradation of $\osp(1|2n)$~\cite{STV}.
We consider here a class of infinite-dimensional representations of $\osp(1|2n)$, namely the ones with
highest weight given by $(-\frac{p}{2},-\frac{p}{2},\ldots, -\frac{p}{2})$ in the $\delta$-basis.
For this representation, the Dynkin labels are $[0,0,\ldots,0,-p]$. 
The structure and character of this representation have been determined in~\cite{paraboson}. 
Using the notation $x_i=\hbox{e}^{-\delta_i}$, one has:
\begin{equation}
\ch [0,0,\ldots,0,-p]_{\mathfrak{osp}(1|2n)} = (x_1\cdots x_n)^{p/2} \sum_{\lambda,\ \ell(\lambda)\leq p} s_\lambda(x).
\label{char-osp-1}
\end{equation}
This is an infinite sum over all partitions of length at most $p$. Since $s_\lambda(x)=0$ if $\ell(\lambda)>n$, the sum is actually over all partitions satisfying $\ell(\lambda)\leq\min(n,p)$.
Applying the above specialization $F_0$, one finds:
\begin{equation}
\dim_t [0,0,\ldots,0,-p]_{\mathfrak{osp}(1|2n)} = \sum_{\lambda,\ \ell(\lambda)\leq\min(n,p)} \dim V_{\mathfrak{gl}(n)}^{\lambda} t^{|\lambda|}.
\label{tdim-osp1}
\end{equation}
This infinite sum can be rewritten in an alternative form, see~\cite{STV}.
Some examples for $\osp(1|6)$ are given by:
\begin{align*}
\dim_t [0,0,-1]_{\mathfrak{osp}(1|6)}&= \frac{1-3t^2+3t^4-t^6}{(1-t)^3(1-t^2)^3}=\frac{1}{(1-t)^3} \\
&= 1+3t+6t^2+10t^3+15t^4+\cdots\\
\dim_t [0,0,-2]_{\mathfrak{osp}(1|6)}&= \frac{1-t^3}{(1-t)^3(1-t^2)^3}
= 1+3t+9t^2+18t^3+36t^4+\cdots\\
\dim_t [0,0,-3]_{\mathfrak{osp}(1|6)}&= \frac{1}{(1-t)^3(1-t^2)^3}= 1+3t+9t^2+19t^3+39t^4+\cdots
\end{align*}

\section{Superdimensions for $\osp(2m+1|2n)$ and $\osp(2m|2n)$}
\label{sec:4}

In this section we mainly summarize some of the main results of~\cite{STV}.
For the Lie superalgebra $B(m,n)=\mathfrak{osp}(2m+1|2n)$, we work with the distinguished set
of simple roots in the $\epsilon$-$\delta$-basis~\cite{Kac,dict} 
\begin{equation}
\delta_1-\delta_2,\  \ldots, \delta_{n-1}-\delta_n,\ \delta_n-\epsilon_1,\ \epsilon_1-\epsilon_2,\ \ldots, \epsilon_{m-1}-\epsilon_m,\ \epsilon_m. 
\label{osp2m12n}
\end{equation}
The relevant $\gl(m|n)$ subalgebra is spanned by the root vectors corresponding to $\delta_i-\delta_j$, $\epsilon_i-\epsilon_j$, $\pm(\epsilon_i-\delta_j)$,
and $\g=\osp(2m+1|2n)$ admits a $\mathbb{Z}$-gradation 
$\mathfrak{g}=\mathfrak{g}_{-2}\oplus\mathfrak{g}_{-1}\oplus\mathfrak{g}_0\oplus\mathfrak{g}_{+1}\oplus\mathfrak{g}_{+2}$
with $\g_0=\gl(m|n)$.

The class of representations to be considered are the irreducible highest weight representations with highest weight given by
$(\frac{p}{2},\ldots,\frac{p}{2};-\frac{p}{2},\ldots, -\frac{p}{2})$ in the $\epsilon$-$\delta$-basis.
This representation has Dynkin labels $[0,0,\ldots,0,p]$.
Using $x_i=\hbox{e}^{-\epsilon_i}$, $y_i=\hbox{e}^{-\delta_i}$, the following character formula holds~\cite{parast,STV}:
\begin{equation}
\ch [0,\ldots,0,p]_{\mathfrak{osp}(2m+1|2n)} = (y_1\cdots y_n/x_1\cdots x_m)^{p/2} \sum_{\lambda,\ \lambda_1\leq p} s_\lambda(x|y).
\label{char-Bmn}
\end{equation}
Here the sum is over all partitions $\lambda$ inside the $(m,n)$-hook (otherwise $s_\lambda(x|y)$ is zero anyway) with 
$\lambda_1\leq p$, or equivalently $\ell(\lambda')\leq p$.
Applying $F_1$, one should (apart from the factor in front of the above
sum) specify $x_i=t$ and $y_j=-t$ in the above character, and so one finds
\begin{align}
\sdim_t [0,\ldots,0,p]_{\mathfrak{osp}(2m+1|2n)} &= \sum_{\lambda,\ \lambda_1\leq p} s_\lambda(t,\ldots,t|-t,\ldots,-t) \nonumber\\
&= \sum_{\lambda,\ \lambda_1\leq p} s_\lambda(1,\ldots,1|-1,\ldots,-1)\, t^{|\lambda|}  \nonumber\\
&= \sum_{\lambda,\ \lambda_1\leq p} \sdim V_{\mathfrak{gl}(m|n)}^{\lambda} \, t^{|\lambda|}. 
\label{sdim-Bmn}
\end{align}
But superdimension formulas for covariant representations of $\mathfrak{gl}(m|n)$ are well known~\cite{King1983}, 
and reduce to dimensions of $\gl(k)$ irreps: 
\begin{equation}
\sdim V_{\mathfrak{gl}(n+k|n)}^{\lambda} =\dim V_{\mathfrak{gl}(k)}^{\lambda}, \qquad
\sdim V_{\mathfrak{gl}(m|m+k)}^{\lambda} =(-1)^{|\lambda|}\dim V_{\mathfrak{gl}(k)}^{\lambda'}.
\label{dim-sdim}
\end{equation}
In particular, when $m=n$, $\sdim V_{\mathfrak{gl}(n|n)}^{\lambda} =0$ unless $\lambda$ is the zero partition $(0)$.
Note that~\eqref{dim-sdim} implies: when $\ell(\lambda)>k$ then $\sdim V_{\mathfrak{gl}(n+k|n)}^{\lambda}=0$;
when $\lambda_1>k$ then $\sdim V_{\mathfrak{gl}(m|m+k)}^{\lambda}=0$.
Applying this to~\eqref{sdim-Bmn} leads to three cases.

\vskip 1mm
\noindent {\bf Case 1: $m=n$, $\mathfrak{osp}(2n+1|2n)$.} 
All superdimensions of covariant representations of $\mathfrak{gl}(n|n)$ are zero, except when $\lambda=(0)$. Hence:
\begin{equation}
\sdim_t [0,\ldots,0,p]_{\mathfrak{osp}(2n+1|2n)} = 1.
\end{equation}

\vskip 1mm
\noindent {\bf Case 2: $m=n+k$, $\mathfrak{osp}(2n+2k+1|2n)$.}
This is the most interesting case. The infinite sum in~\eqref{sdim-Bmn} reduces to a finite sum:
\begin{align}
\sdim_t [0,\ldots,0,p]_{\mathfrak{osp}(2m+1|2n)} &= \sum_{\lambda,\ \lambda_1\leq p} \dim V_{\mathfrak{gl}(k)}^{\lambda} \, t^{|\lambda|} \nn\\
&= \sum_{\lambda,\ \lambda_1\leq p,\ \ell(\lambda)\leq k} \dim V_{\mathfrak{gl}(k)}^{\lambda} \, t^{|\lambda|}.
\label{sdim-Bk1}
\end{align}
This coincides with example~\eqref{e5}. Hence we can write
\begin{equation}
\sdim_t [0,0,\ldots,0,p]_{\mathfrak{osp}(2n+2k+1|2n)} = \dim_t [0,\ldots,0,p]_{\mathfrak{so}(2k+1)}.
\label{sdim-Bk3}
\end{equation}

\vskip 1mm
\noindent {\bf Case 3: $n=m+k$, $\mathfrak{osp}(2m+1|2m+2k)$.} One finds:
\begin{align}
\sdim_t [0,\ldots,0,p]_{\mathfrak{osp}(2m+1|2n)} &= \sum_{\lambda,\ \lambda_1\leq p,\ \lambda_1\leq k}
(-1)^{|\lambda|} \dim V_{\mathfrak{gl}(k)}^{\lambda'} \, t^{|\lambda|} \nn\\
&= \sum_{\mu,\ \ell(\mu)\leq\min(p,k)} \dim V_{\mathfrak{gl}(k)}^{\mu} \, (-t)^{|\mu|}.
\label{sdim-sBk1}
\end{align}
The right hand side is the same expression as~\eqref{tdim-osp1}, so 
\begin{equation}
\sdim_t [0,0,\ldots,0,p]_{\mathfrak{osp}(2m+1|2m+2k)} = \dim_{-t} [0,\ldots,0,-p]_{\mathfrak{osp}(1|2k)}.
\label{sdim-sBk3}
\end{equation}

So in all three cases, the superdimension for $\mathfrak{osp}(2m+1|2n)$ simplifies and reduces 
to a dimension of $\mathfrak{so}(2m+1-2n)$ or $\mathfrak{osp}(1|2n-2m)$.

Let us now turn to the Lie superalgebra $D(m,n)=\mathfrak{osp}(2m|2n)$. 
The distinguished set of simple roots in the $\epsilon$-$\delta$-basis is
\begin{equation}
\delta_1-\delta_2,\  \ldots, \delta_{n-1}-\delta_n,\ \delta_n-\epsilon_1,\ \epsilon_1-\epsilon_2,\ \ldots, \epsilon_{m-2}-\epsilon_{m-1},\ \epsilon_{m-1}-\epsilon_m,\ \epsilon_{m-1}+\epsilon_m. 
\label{osp2m2n}
\end{equation}
It will be helpful to see this superalgebra in the subalgebra chain $\osp(2m+1|2n)\supset\osp(2m|2n)\supset\gl(m|n)$.

For the irreducible highest weight representation of $\mathfrak{osp}(2m|2n)$ with highest weight given by
$(\frac{p}{2},\ldots,\frac{p}{2};-\frac{p}{2},\ldots, -\frac{p}{2})$, 
with Dynkin labels $[0,0,\ldots,0,p]$, the character was determined in~\cite{STV}:
\begin{equation}
\ch [0,\ldots,0,p]_{\mathfrak{osp}(2m|2n)} = (y_1\cdots y_n/x_1\cdots x_m)^{p/2} 
\sum_{\lambda\in{\cal B},\ \lambda_1\leq p} s_\lambda(x|y).
\label{char-Dmn}
\end{equation}
Herein, ${\cal B}$ denotes the set of partitions for which each part appears twice (including the zero partition).
Thus, one finds
\begin{equation}
\sdim_t [0,\ldots,0,p]_{\mathfrak{osp}(2m|2n)} = 
\sum_{\lambda\in{\cal B},\ \lambda_1\leq p} \sdim V_{\mathfrak{gl}(m|n)}^{\lambda} \, t^{|\lambda|}. 
\label{sdim-Dmn}
\end{equation}
This expression allows once again to deduce superdimension formulas in three cases: $m=n$, $m>n$ and $m<n$, see~\cite{STV}.
Let us give here the formula for $m>n$, i.e.\ $m=n+k$, or $\mathfrak{osp}(2n+2k|2n)$.
From~\eqref{sdim-Dmn} one has:
\begin{align}
\sdim_t [0,\ldots,0,p]_{\mathfrak{osp}(2m|2n)} &= \sum_{\lambda\in{\cal B},\ \lambda_1\leq p} \dim V_{\mathfrak{gl}(k)}^{\lambda} \, t^{|\lambda|} \nn\\
&= \sum_{\lambda\in{\cal B},\ \lambda_1\leq p,\ \ell(\lambda)\leq k} \dim V_{\mathfrak{gl}(k)}^{\lambda} \, t^{|\lambda|}.
\label{sdim-Dk1}
\end{align}
This is to be compared to known characters of $\so(2k)$ irreps~\cite{STV}, where a distinction should be made between $k$ even and $k$ odd.
For $k$ even, one has
\begin{equation}
\ch [0,\ldots,0,p]_{\mathfrak{so}(2k)} = (x_1\cdots x_k)^{-p/2} \sum_{\lambda \in {\cal B};\ \lambda_1\leq p,\; \ell(\lambda)\leq k} s_\lambda (x).
\label{char25}
\end{equation}
For $k$ odd,
\begin{equation}
\ch [0,\ldots,p,0]_{\mathfrak{so}(2k)} = (x_1\cdots x_k)^{-p/2} \sum_{\lambda \in {\cal B}:\ \lambda_1\leq p,\; \ell(\lambda)\leq k-1} s_{\lambda} (x).
\label{char25b}
\end{equation}
Comparing with \eqref{sdim-Dk1}, yields:
\begin{equation}
\sdim_t [0,0,\ldots,0,p]_{\mathfrak{osp}(2n+2k|2n)} = \left\{
\begin{array}{ll}
\displaystyle \dim_t [0,\ldots,0,0,p]_{\mathfrak{so}(2k)} & \hbox{ for }k\hbox{ even}, \\
\displaystyle \dim_t [0,\ldots,0,p,0]_{\mathfrak{so}(2k)} & \hbox{ for }k\hbox{ odd}. 
\end{array} \right.
\label{sdim-Dk3}
\end{equation}
Here, the convention for the order of the simple roots of $\mathfrak{so}(2k)$ is $\epsilon_1-\epsilon_2,
\ldots, \epsilon_{k-1}-\epsilon_k,\epsilon_{k-1}+\epsilon_k$.

\section{Characters of ``fork'' representations for $\so(2m)$ and $\osp(2m|2n)$}
\label{sec:5}

\begin{figure}[htb]
\sidecaption
\includegraphics[scale=.8]{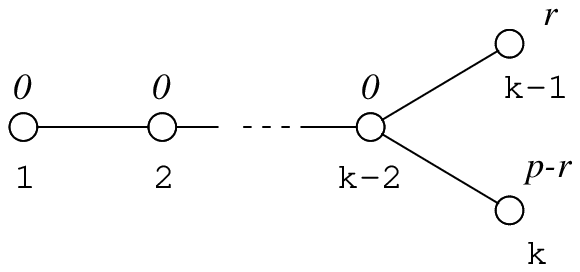}
%
%
\caption{Dynkin diagram of the fork representation of $\so(2k)$ \qquad\qquad\qquad\qquad\qquad\qquad\qquad}
\label{fig:1}       
\end{figure}

The characters of $\so(2k)$ and $\so(2k+1)$, used in the previous section, should be seen in the context of 
the subalgebra chain $\so(2k+1)\supset\so(2k)\supset\gl(k)$.
In~\eqref{e4} we obtained
\begin{equation}
\ch [0,\ldots,0,p]_{\mathfrak{so}(2k+1)} = (x_1\cdots x_k)^{-p/2} \sum_{\lambda_1\leq p,\; \ell(\lambda)\leq k} s_\lambda (x).
\end{equation}
Essentially, this is the branching $\mathfrak{so}(2k+1) \supset \mathfrak{gl}(k)$, since Schur functions are characters of $\gl(k)$ irreps. 
Considering the representation with respect to the branching $\so(2k+1)\supset\so(2k)$, one finds (using Weyl's character formula):
\begin{equation}
\ch [0,\ldots,0,p]_{\mathfrak{so}(2k+1)} = \sum_{r=0}^p \ch [0,\ldots,r,p-r]_{\mathfrak{so}(2k)}.
\label{e28}
\end{equation}
The $\so(2k)$ representations with Dynkin labels $[0,\ldots,r,p-r]$ are sometimes referred to as fork representations, since the only non-zero
Dynkin labels appear at the fork nodes of the diagram, see Fig.~\ref{fig:1}

\begin{figure}[htb]
\sidecaption
\includegraphics[scale=.8]{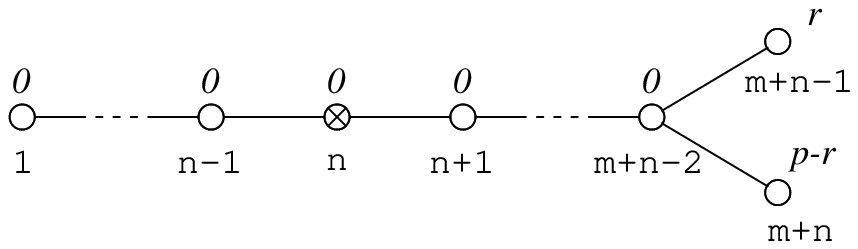}
%
%
\caption{Dynkin diagram of the fork representation of $\osp(2m|2n)$ \qquad\qquad\qquad\qquad\qquad\qquad\qquad}
\label{fig:2}       
\end{figure}

The $\mathfrak{so}(2k)$ characters -- in terms of Schur functions -- that were used in the identification of the right hand side of~\eqref{sdim-Dk1}
were for the representations $[0,\ldots,0,p]$ and $[0,\ldots,0,p,0]$.
Given \eqref{e28}, the question is how to write the character of the other $\mathfrak{so}(2k)$ 
fork representations $[0,\ldots,r,p-r]$ as a sum of Schur functions?
Or in other words, what is the branching $\mathfrak{so}(2k) \supset \mathfrak{gl}(k)$ for these representations?
The answer is given by:

\begin{theorem}
For $k$ even, one has
\begin{equation}
\ch [0,\ldots,0,r,p-r]_{\mathfrak{so}(2k)} = (x_1\cdots x_k)^{-p/2} 
\sum_{\lambda_1\leq p,\; \ell(\lambda)\leq k;\; \lambda\in{\cal B}_r} s_\lambda (x).
\label{char-so1}
\end{equation}
Herein, ${\cal B}_r$ stands for the set of partitions of ${\cal B}$ to which a horizontal strip of length~$r$ is attached.
(Recall that ${\cal B}$ is the set of partitions for which each part appears twice.)
The first condition ($\lambda_1\leq p,\; \ell(\lambda)\leq k$) means that (the Young diagram of) $\lambda$ fits inside the $k\times p$ rectangle.
Similarly, for $k$ odd:
\begin{equation}
\ch [0,\ldots,0,r,p-r]_{\mathfrak{so}(2k)} = (x_1\cdots x_k)^{-p/2} 
\sum_{\lambda_1\leq p,\; \ell(\lambda)\leq k;\; \lambda\in{\cal B}_{p-r}} s_\lambda (x).
\label{char-so2}
\end{equation}
\end{theorem}

The proof is technical and can be obtained using the branching rules for $\mathfrak{so}(2k) \supset \mathfrak{gl}(k)$ described in~\cite{KW}.
Note that, in accordance with~\eqref{e28}, the union of all partitions of ${\cal B}_r$ in the $k\times p$ rectangle, for $r=0,1,\ldots,p$, 
is equal to the set of all partitions in the rectangle.

In order to illustrate the sets ${\cal B}_r$, let us give some examples for $\so(8)$.
\begin{align*}
\ch [0,0,0,1]_{\mathfrak{so}(8)} &= (x_1\cdots x_4)^{-1/2}(1+s_{(1,1)}+s_{(1,1,1,1 )})\\
\ch [0,0,1,0]_{\mathfrak{so}(8)} &= (x_1\cdots x_4)^{-1/2}(s_{(1)}+s_{(1,1,1)})\\
\ch [0,0,0,2]_{\mathfrak{so}(8)} &= (x_1\cdots x_4)^{-1}(1+s_{(1,1)}+s_{(2,2)}+s_{(1,1,1,1)}+s_{(2,2,1,1)}\\
& +s_{(2,2,2,2)})\\
\ch [0,0,1,1]_{\mathfrak{so}(8)} &= (x_1\cdots x_4)^{-1}(s_{(1)}+s_{(2,1)}+s_{(1,1,1)}+s_{(2,2,1)}+s_{(2,1,1,1)}\\
& +s_{(2,2,2,1)})\\
\ch [0,0,2,0]_{\mathfrak{so}(8)} &= (x_1\cdots x_4)^{-1}(s_{(2)}+s_{(2,1,1)}+s_{(2,2,2)})\\
\ch [0,0,0,3]_{\mathfrak{so}(8)} &= (x_1\cdots x_4)^{-3/2}(1+s_{(1,1)}+s_{(2,2)}+s_{(1,1,1,1)}+s_{(3,3)}\\
& +s_{(2,2,1,1)}+s_{(3,3,1,1)}+s_{(2,2,2,2)}+s_{(3,3,2,2)}+s_{(3,3,3,3)})\\
\ch [0,0,1,2]_{\mathfrak{so}(8)} &= (x_1\cdots x_4)^{-3/2}(s_{(1)}+s_{(2,1)}+s_{(1,1,1)}+s_{(2,1,1,1)}+s_{(2,2,1)}\\
&+s_{(3,2)}+s_{(2,2,2,1)}+s_{(3,2,1,1)}+s_{(3,3,1)}+s_{(3,2,2,2)}\\
&+s_{(3,3,2,1)}+s_{(3,3,3,2)})\\
\ch [0,0,2,1]_{\mathfrak{so}(8)} &= (x_1\cdots x_4)^{-3/2}(s_{(2)}+s_{(2,1,1)}+s_{(3,1)}+s_{(2,2,2)}+s_{(3,1,1,1)}\\
&+s_{(3,2,1)}+s_{(3,2,2,1)}+s_{(3,3,2)}+s_{(3,3,3,1)})\\
\ch [0,0,3,0]_{\mathfrak{so}(8)} &= (x_1\cdots x_4)^{-3/2}(s_{(3)}+s_{(3,1,1)}+s_{(3,2,2)}+s_{(3,3,3)})
\end{align*}
From these examples, one can indeed see that for representations $[0,0,0,p]$, only partitions appear for which each part is repeated twice (inside the 
$4\times p$ rectangle).
The partitions appearing in, e.g., $[0,0,2,1]$ are obtained from those of $[0,0,0,3]$ by attaching a horizontal strip of length 2.
Note that indeed the union of all partitions appearing in, e.g., $[0,0,0,3]$, $[0,0,1,2]$, $[0,0,2,1]$ and $[0,0,3,0]$ give indeed
{\em all} partitions inside the $4\times 3$ rectangle.

But now we can extend the analogy that we observed between representations $[0,\ldots,0,p]$ of $\mathfrak{osp}(m|n)$ 
and the corresponding ones of $\mathfrak{so}(m-n)$. 
For $\osp(2m+1|2n)$, one should compare equation~\eqref{char-Bmn} with~\eqref{e5}.
For $\osp(2m|2n)$, one should compare~\eqref{char-Dmn} with~\eqref{char25}.
For all these cases, the character of the corresponding representation (expressed in terms of Schur functions) is the same, 
up to the extra condition $\ell(\lambda)\leq k$ for $\so(2k)$.
We conjecture that this correspondence also holds for the characters of fork representations of $\osp(2m|2n)$ (see Fig.~\ref{fig:2}),
by dropping the condition $\ell(\lambda)\leq k$  in~\eqref{char-so1}.

\begin{conjecture}
For $|m-n|$ even, one has
\begin{equation}
\ch [0,\ldots,0,r,p-r]_{\mathfrak{osp}(2m|2n)} = (y_1\cdots y_n/x_1\cdots x_m)^{p/2} 
\sum_{\lambda_1\leq p,\; \lambda\in{\cal B}_r} s_\lambda (x/y).
\label{conj}
\end{equation}
So in this case we have an expansion as an infinite sum of supersymmetric Schur functions,
labeled by partitions $\lambda$ inside the $(m,n)$-hook, of width at most $p$, and belonging to ${\cal B}_r$.
\end{conjecture}

For $|m-n|$ odd, the result is similar, with ${\cal B}_r$ replaced by ${\cal B}_{p-r}$, following~\eqref{char-so2}.

Note that this conjecture also has some interesting consequences, and yields the equivalence of~\eqref{e28}:
\begin{equation}
\ch [0,\ldots,0,p]_{\mathfrak{so}(2m+1|2n)} = \sum_{r=0}^p \ch [0,\ldots,r,p-r]_{\mathfrak{so}(2m|2n)}.
\label{e28b}
\end{equation}
Indeed, the expansion of the left hand side is given by~\eqref{char-Bmn}, and involves all partitions $\lambda$ with $\lambda_1\leq p$.
The expansion of the terms in the right hand side is given by~\eqref{conj}; each term involves the partitions of ${\cal B}_r$ with $\lambda_1\leq p$.
Clearly, $\{\lambda \;|\; \lambda_1\leq p\}$ is the disjoint union of the sets
\begin{equation}
\{ \lambda \in {\cal B}_r \;|\; \lambda_1\leq p\}, \qquad r=0,1,\ldots,p.
\label{Br}
\end{equation}
Obviously, every element of \eqref{Br} belongs to $\{\lambda \;|\; \lambda_1\leq p\}$.
The other way round, when $\lambda$ is an arbitrary partition with $\lambda_1\leq p$, one should make the following construction.
For $\lambda=(\lambda_1,\lambda_2,\lambda_3,\lambda_4, \ldots)$, let $\mu_1=\mu_2=\lambda_2$, $\mu_3=\mu_4=\lambda_4$, etc.;
thus $\mu\in {\cal B}$ (all parts appear twice). And $\lambda-\mu$ is by construction a horizontal strip of length $r=|\lambda|-|\mu|$,
where $r\leq p$ since $\lambda_1\leq p$.
So $\lambda$ belongs to a unique set of~\eqref{Br} for some $r\in\{0,1,\ldots,p\}$.
Now~\eqref{e28b} follows.

To conclude, in the current paper we have first analyzed characters and superdimensions for representations of the form $[0,\ldots,0,p]$ 
for $\mathfrak{osp}(2m+1|2n)$ and $\osp(2m|2n)$, and related them to characters and dimensions of $\so(2k+1)$ and $\so(2k)$ (for $k=m-n$).
Exploiting this correspondence, we conjecture that it also holds for fork representations of the form $[0,\ldots,0,r,p-r]$ for $\mathfrak{osp}(2m|2n)$.
For this purpose, we have deduced characters of the corresponding fork representations of $\so(2k)$.
The formal proof of the conjecture might be difficult or technical. 
One way is to try and use characters of more general $\mathfrak{osp}(m|n)$ tensors which were studied in~\cite{Cheng}. 
Here, the character formulas correspond to alternating series of $S$-functions, which are not easy to handle.
Another way is to make use of the explicit construction of the representation $[0,\ldots,0,p]_{\mathfrak{so}(2m+1|2n)}$ 
in~\cite{parast}.
This method is in principle straightforward, but might be difficult to perform because of the complicated matrix elements appearing for these representations.

\begin{acknowledgement}
NIS and JVdJ were supported by the Joint Research Project ``Lie superalgebras - applications in quantum theory'' in the framework of an international collaboration programme between the Research Foundation -- Flanders (FWO) and the Bulgarian Academy of Sciences. 
NIS was partially supported by the Bulgarian National Science Fund, grant DN~18/1.
This research (JT-M) was supported in part by the Intramural Research Program of the NIH, U.S. National Library of Medicine.
\end{acknowledgement}
\end{document}